# Frequency Dependence of Polarization of Zebra Pattern in Type-IV Solar Radio Bursts

Short title: Frequency-dependent zebra-pattern polarization

Kazutaka Kaneda[1], H. Misawa[1], K. Iwai[2], F. Tsuchiya[1], and T. Obara[1]

[1]Planetary Plasma and Atmospheric Research Center, Tohoku University, Sendai, Miyagi 980-8578, Japan; k.kaneda@pparc.gp.tohoku.ac.jp

[2]National Institute of Information and Communications Technology, 4-2-1, Nukui-Kitamachi, Koganei, Tokyo 184-8795, Japan

## Abstract

We investigated the polarization characteristics of a zebra pattern (ZP) in a type-IV solar radio burst observed with AMATERAS on 2011 June 21 for the purpose of evaluating the generation processes of ZPs. Analyzing highly resolved spectral and polarization data revealed the frequency dependence of the degree of circular polarization and the delay between two polarized components for the first time. The degree of circular polarization was 50-70 % right-handed and it varied little as a function of frequency. Cross-correlation analysis determined that the left-handed circularly polarized component was delayed by 50–70 ms relative to the right-handed component over the entire frequency range of the ZP and this delay increased with the frequency. We examined the obtained polarization characteristics by using pre-existing ZP models and concluded that the ZP was generated by the double-plasma-resonance process. Our results suggest that the ZP emission was originally generated in a completely polarized state in the O-mode and was partly converted into the X-mode near the source. Subsequently, the difference between the group velocities of the O-mode and X-mode caused the temporal delay.

## 1. INTRODUCTION

Type-IV solar radio bursts are broadband emissions observed in the metric- to decimetric-wavelength range. They are generally emitted in association with a large flare and considered to be generated by non-thermal electrons trapped in a closed magnetic structure (Mclean & Labrum 1985). Past observations have revealed various complex spectral fine structures in type-IV bursts, such as fiber bursts, broadband pulsations, and zebra patterns (Kuijpers 1980). Because these fine structures are thought to be generated by micro-scale physical processes of plasma waves and energetic particles during wave generation and/or propagation, their spectral characteristics should reflect the plasma conditions of the source region and the plasma processes occurring in the corona.

The zebra pattern (ZP) is a characteristic spectral structure with a number of nearly parallel drifting narrowband stripes of enhanced emission. The number of stripes typically ranges from 5 to 20 (70 at most), and the frequency separation between adjacent emission bands increases with an increase of frequency (Chernov 2006).

To explain ZPs, more than ten theoretical models have been proposed and there is no widely accepted interpretation. Most of these theories are based on the plasma emission mechanism, which involves the excitation of electrostatic waves and their conversion into electromagnetic waves (e.g. Rosenberg 1972; Chernov 1990; Ledenev et al. 2006).

The most popular model is the so-called double plasma resonance (DPR) model, proposed by Zheleznyakov & Zlotnik (1975), who suggested that ZP stripes are generated in a distributed source along the field line. In the framework of this model, the plasma waves are excited at certain levels, where the upper-hybrid frequency $f_{UH}$ coincides with the harmonics of the electron cyclotron frequency $f_c$:

$$f_{UH} = \sqrt{f_p^2 + f_c^2} = sf_c, \qquad (1)$$

where $f_p$ is the electron plasma frequency and $s$ is the harmonic number.

The generation processes of ZPs influence not only their spectral appearance but also their polarization characteristics. Some studies on the polarization of ZPs have been reported (e.g., Chernov et al. 1975, Chernov & Zlobec 1995, Kuznetsov 2008, and Zlotnik et al. 2014). In most ZP events observed until now, the ZP was strongly polarized in the same sense as the background continuum and the degree of circular polarization ranged between 20–100 %. The wave mode of ZPs is considered to be the O-mode. The wave mode is determined from the sense of rotation of the electric vector of the wave and the magnetic polarity in the source region, assuming that the emission was generated just above the leading sunspot of the active region (Chernov 2006). Using theoretical calculations, Zlotnik et al. (2014) also concluded that a ZP is emitted at a fundamental plasma frequency and polarized in the O-mode; this was supported by spatially resolved observations of ZPs (Chen et al. 2011). For several events, the delay between the right- and left-handed circularly polarized (RCP and LCP) components was examined. This delay was interpreted by the difference in the group velocities of O- and X-modes (Chernov & Zlobec 1995).

However, the existing ZP observations are limited in the fixed frequency band with a low frequency resolution. In order to reveal the detailed polarization characteristics of such a small-scale phenomenon, broadband observation with high-frequency resolution that can resolve each stripe of the ZP is necessary. Furthermore, to determine how the characteristics of the emission change according to the height in the corona, we must determine how the polarization characteristics of the emission depend on the frequency.

In this study, one ZP event was investigated for the purpose of evaluating the generation and propagation processes proposed in previous studies. We analyzed the highly resolved spectrum and polarization data and determined the polarization characteristics and their frequency dependence. A description of the instrumentation and an overview of the ZP event are provided in Section 2. In Section 3, the data analysis and the results are described. We discuss the results in Section 4 and present a summary of the paper in Section 5.

## 2. OBSERVATIONS

AMATERAS (the Assembly of Metric-band Aperture Telescope and Real-time Analysis System) is a solar radio telescope dedicated to spectropolarimetry in the metric range (Iwai et al. 2012). It can observe simultaneously the RCP and LCP components of solar radio bursts in the frequency range 150–500 MHz with a time resolution of 10 ms and a frequency resolution of 61 kHz. Hence, it can be used to acquire data for the analysis of fine spectrum structures and the polarization characteristics of solar radio bursts (e.g., Iwai et al. 2013, 2014, Katoh et al. 2014).

On 2011 June 21, a type-IV burst accompanied by a clear ZP was observed with

AMATERAS. This type-IV burst was emitted in association with the C7.7 flare. Figure 1 shows the GOES X-ray flux (top panel) and the dynamic spectrum of the RCP component of the type-IV bursts observed with AMATERAS. The flare lasted about 3 h; it began at 01:22 UT, peaked at 03:25 UT, and ended at approximately 04:30 UT. Figure 2(a) shows the full disk image at 131 Å observed with SDO/AIA. A bright active region (AR11236) was located near the disk center. The HMI magnetogram corresponding to the white square in Figure 2(a) is shown in Figure 2(b). The active region was bipolar and consisted of a leading spot with negative polarity and a following spot with positive polarity. The green contours in Figure 2(b) show the radio brightness temperature at 17 GHz observed by the Nobeyama Radioheliograph (NoRH; Nakajima et al. 1994) at the peak time. This emission is considered non-thermal gyro-synchrotron emission. The brightest point at 17 GHz was located above the leading spot. The time evolution of AR11236 in the 131 Å line is shown in Figure 2(c)–(e). In association with the flare brightening in the GOES X-ray flux, the brightening of magnetic loop structures around the active region can be clearly seen. Since there was only one flare before and after the bursts, the bursts probably emanated from this active region.

During the flare, several type-IV bursts occurred intermittently and the ZPs were detected in only one burst, at around 03:20 UT, which is indicated by the red dashed rectangle in Figure 1. Figure 3 shows the dynamic spectrum of the type-IV burst in which the ZP appeared. The ZPs are denoted by the red dashed rectangles (around 03:20 UT, 03:22 UT and 03:24 UT). The time of appearance of the ZPs approximately coincided with the peak time of the flare in the GOES X-ray flux and the brightening in EUV 131 Å (see Figure 2(c)-(e)). In addition, at the same time as the appearance of the ZPs, similar continuum bursts were observed in the lower frequency range (25–180 MHz) with the Learmonth Spectrograph in Australia. The detection of similar phenomenons at two different observatories and the correspondence in time between the flare brightening and bursts indicate the solar origin of the observed ZPs.

Figure 4 shows a magnified image of the strongest part of the ZP (03:22:06–03:22:36 UT). The ZP stripes are evident in the frequency range 160–210 MHz and are remarkably enhanced in fast drifting envelopes, such as type-III bursts or broadband pulsations, which are indicated by white dotted circles. Although the ZP stripes are enhanced and attenuated synchronously with the fast drifting envelopes, all stripes seem to be connected continuously with each other over the examined time range. Therefore, we considered a series of ZP stripes as one event. The total number of stripes was approximately 30 and the frequency separation between them increased with frequency (1.4 MHz at 160 MHz and 2.5 MHz at 200 MHz). These parameters are comparable to those of previous studies (e.g., Chernov 2006) and consistent with the DPR theory.

## 3. ANALYSIS AND RESULTS

ZP stripes can be observed in both polarized components, as shown in Figure 4. The intensity of the emission was stronger in the RCP than in the LCP component. The degree of circular polarization $p$ was calculated from the radio intensity of the RCP, $I_R$, and LCP, $I_L$, components using the following equation

$$p = \frac{I_R - I_L}{I_R + I_L} \times 100 \ (\%). \tag{2}$$

In the examined event, the degree of right-handed circular polarization was 50–70 % and this value was almost constant throughout the entire time and frequency range shown in the middle panel of Figure 5. The polarization of the ZP stripes was stronger than the background continuum by approximately 20%. The calculated degree of circular polarization does not represent the polarization of the ZP stripes themselves, because it also contains the polarization of the background continuum emission.

The temporal delay of the LCP (red line) relative to the RCP (black line) component was detected, as shown in the top panel of Figure 5. The delay was evaluated by cross-correlation analysis (e.g., Benz & Pianezzi 1997; Kuznetsov 2008) using the following steps. First, the cross-correlation coefficient $C(l)$ of the RCP and LCP

components was calculated in a selected time interval at a certain frequency channel according to

$$C(l) = \frac{\sum_{i=1}^{n-l} I_R(t_i) I_L(t_{i+l})}{\sqrt{\sum_{i=1}^{n} I_R(t_i)^2 \sum_{i=1}^{n} I_L(t_i)^2}}. \tag{3}$$

The summation in the numerator was calculated over all possible pairs for the lag $l$, and $n$ is the total number of data points in the selected time interval. Then, the calculated function was interpolated by cubic spline functions to increase the resolution in the calculation of the delay. The value of the lag for which the maximum cross-correlation coefficient was obtained was determined to be the delay between the RCP and LCP components. The above processes were repeated for all frequency channels in which the ZP appeared. Subsequently, we obtained the delay over the entire frequency range of the ZP. The error in the determination of the delay $\sigma_{\Delta t}$ was calculated using the law of error propagation, expressed as follows

$$\sigma_{\Delta t} = \frac{\sigma}{C''d} \sqrt{2(1 - C_{max}) \left( \frac{1}{\sum_{i=1}^{n} I_R(t_i)^2} + \frac{1}{\sum_{i=1}^{n} I_L(t_i)^2} \right)}, \tag{4}$$

where $\sigma$ is the standard deviation of the noise signal at the time when the solar activity was quiet (because the values were almost the same in both components, we used that of the RCP component); $C''$ is the second derivative of the interpolated cross-correlation function $C(l)$, calculated at the maximum of the function; $d$ is the temporal resolution of the interpolated function; and $C_{max}$ is the maximum of the cross-correlation function. The accuracy of the delay calculation increases with the intensity of the emission and the length of the selected time interval.

The bottom panel of Figure 5 shows the delay throughout the frequency range of the ZP, as determined by the cross-correlation method. The ordinate represents the lag $l$ for which equation (3) reaches a maximum. In the calculation, we selected the time interval of 7 s from 03:22:22 UT to 03:22:29 UT. The positive delay means that the LCP component followed the RCP one. The red dotted lines show the error $\pm 3\sigma_{\Delta t}$, calculated with equation (4), which is sufficiently small in the frequency range of the ZP enhancement. In the lower frequencies, the error increases because the intensity of the ZP is low. The blue line displays the linear fit in the frequency range 170–200 MHz, where the ZP is clearly visible and the error is relatively small. The frequency ranges of the artificial interferences were ignored in the fitting process. The delay reached 50 ms at 170 MHz and 70 ms at 200 MHz; namely, it increased with an increase of the frequency. The frequency dependence of the delay changes slightly depending on the selected time interval where the cross-correlation is calculated. However, the delay was between 50–70 ms and the frequency dependence was mostly weak during the event.

## 4. DISCUSSION

The wave mode of the ZPs is usually estimated from the sense of rotation of the electric vector of the radio emission and the magnetic polarity of the active region by assuming that the ZP should be emitted above the leading spot (Chernov, 2006). This was supported by spatially resolved observations of ZPs (Chen et al. 2011). In the present study, the magnetic polarity of the leading spot was negative, as shown in Figure 2(b). Therefore, the RCP component corresponds to the O-mode and the LCP component to the X-mode if we assume that the emission originated from just above the leading spot. In fact, non-thermal microwave gyro-synchrotron emission had existed above the leading spot, suggesting that the particle acceleration occurred around that region, although the energy range of electrons and the radio emission mechanism were different from those of the metric bursts (see Figure 2(b)).

Based on the above considerations, the O-mode was dominant in the observed ZP, and the delay of the LCP component indicates a delay of the X-mode waves. These results are consistent with previous studies (Chernov 2006, Chen et al. 2011, and Zlotnik et al. 2014).

The dominance of the O-mode and the increasing frequency separation between adjacent stripes with increasing emission frequency suggest that the observed ZP can be explained by the DPR theory. According to this model (equation (1)), the emission could have been generated as a perfectly polarized state in the O-mode. Because the ZP stripes were evident in both polarized components and their spectral appearance was very similar (see Figure 4), the O-mode and X-mode may have the same origin in the source. Therefore, the observed polarization characteristics suggest that the X-mode possibly resulted from the depolarization effect during propagation and that the delay between the O-mode and X-mode was introduced subsequently.

The most plausible reason for the delay is the difference in the group velocities of the O-mode and the X-mode. Actually, the group velocity of the O-mode waves in magnetized plasma is greater than that of the X-mode waves according to the dispersion relation of the waves. Considering that the observed delay (50–70 ms) and its frequency dependence have been generated by the difference in group velocities, both waves should have propagated through relatively dense plasma with frequencies close to the local plasma frequency. According to commonly used coronal models (e.g., Allen 1947, Dulk & Mclean 1978), the depolarization occurred within tens of thousands kilometers from the source, suggesting that the ZP emission was partly converted into the X-mode near the source. These results imply that the depolarization effect that generated the X-mode should appear near the source of the emission.

The generation process of the X-mode was not confirmed. One of possible generation processes of the X-mode is the effect of mode coupling, proposed by Cohen (1960), during the propagation through a region of a quasi-transverse magnetic field relative to the direction of propagation. However, this process may not be plausible for generating the X-mode. The theories of mode-coupling (e.g., Cohen 1960, Zheleznyakov & Zlotnik 1964) predict that the degree of circular polarization strongly depends on the frequency ($\propto f^{-4}$). In contrast to these theories, the observed degree of circular polarization was almost constant over a wide frequency range, as shown in the middle panel of Figure 5. Therefore, the mode-coupling process may not be plausible, and other related mechanisms such as the mode conversion of the O-mode to the X-mode should be considered. Although in other models similar interpretation can be accepted, the evaluation of them is beyond the scope of this paper.

## 5. SUMMARY

We investigated a ZP event observed with AMATERAS on 2011 June 21 in association with the C7.7 flare. The broadband and high-resolution data from AMATERAS enabled us for the first time to perform a detailed analysis of the polarization and frequency characteristics of ZPs, in contrast to previous observations performed at the fixed frequency band. The ZP appeared in both the RCP and LCP channels and was dominant in the RCP one. The degree of circular polarization was found to be 50–70 % right-handed with almost no frequency dependence. In addition, the delay between the two polarized components was evaluated by cross-correlation analysis. The analysis determined that the LCP was delayed by 50–70 ms relative to the RCP and the delay increased with an increase of the frequency.

From the obtained frequency dependence of the polarization, we evaluated the generation processes of the ZP assuming the DPR model. Our findings suggest that the depolarization from the O-mode to the X-mode occurred near the source of emission and the observed delay was subsequently introduced due to the different group velocities of the two components. Further studies on ZPs should focus on localizing the position where the emission originated. For this purpose, combining interferometric observations with spectral measurements may be an effective strategy.

AMATERAS is a Japanese radio telescope developed and operated by Tohoku University. We are grateful to the GOES and SDO teams for making the data available to us. This work was conducted by a joint research program of the Solar-Terrestrial Environment Laboratory, Nagoya University.

**Figures**

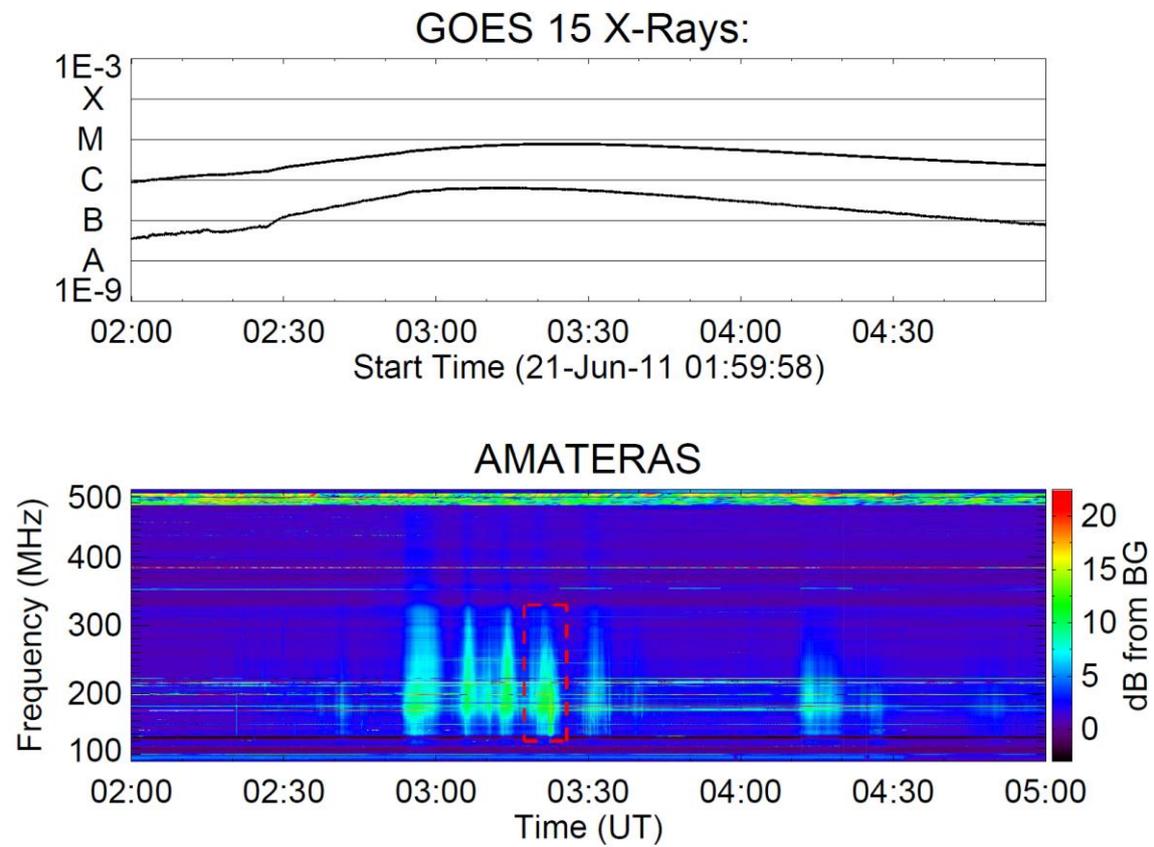

Figure 1

Type-IV burst event observed with AMATERAS on 2011 Jun 21. Top: GOES X-ray flux, bottom: dynamic spectrum of a series of type-IV bursts in RCP. The red dashed rectangle indicates the burst in which the ZP appeared.

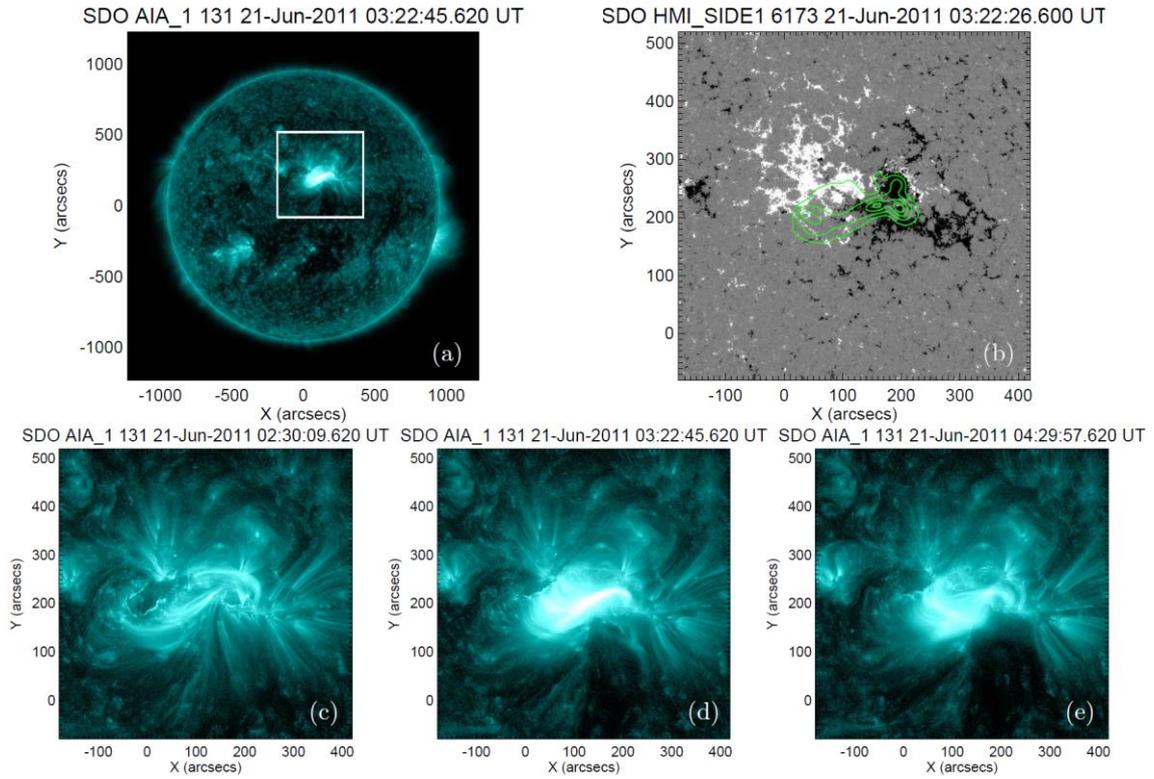

Figure 2

Images of the flare during the bursts observed with SDO/AIA and SDO/HMI. (a): Full disk image in AIA 131 Å. (b): HMI magnetogram of the region indicated by the white square in panel (a), showing AR11236. Green contours show NoRH 17 GHz radio intensity. (c)–(e): Sequence of magnified images of AR11236 in AIA 131 Å.

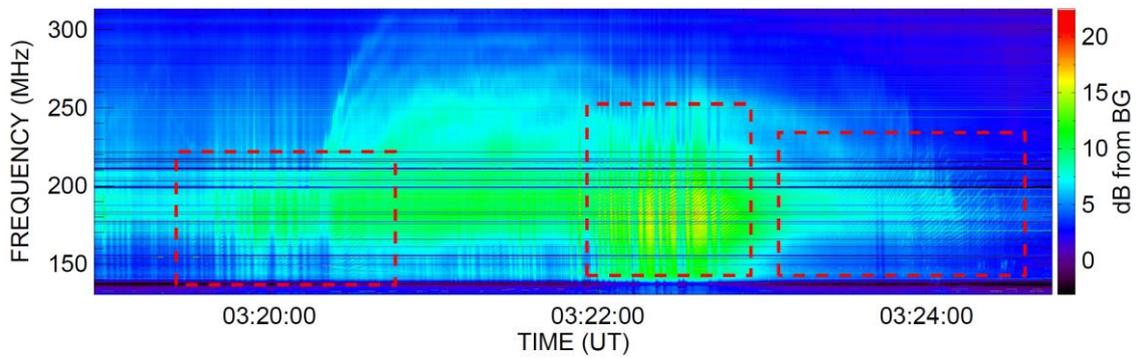

Figure 3

Dynamic spectrum of the RCP component of the type-IV burst observed with AMATERAS on 2011 June 21. The red rectangles show the time periods when the ZPs appeared.

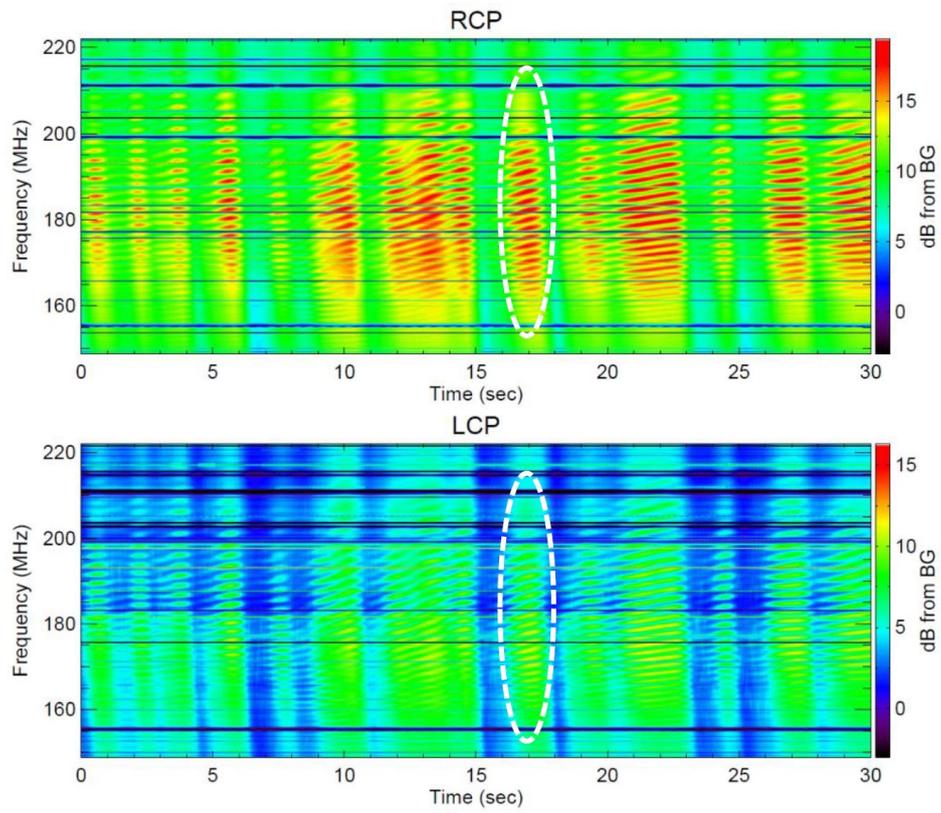

Figure 4

Magnified image of the ZP for 03:22:06–03:22:36 UT (top: RCP, bottom: LCP). The ZP was enhanced in fast drifting envelopes, indicated by the white dotted circles.

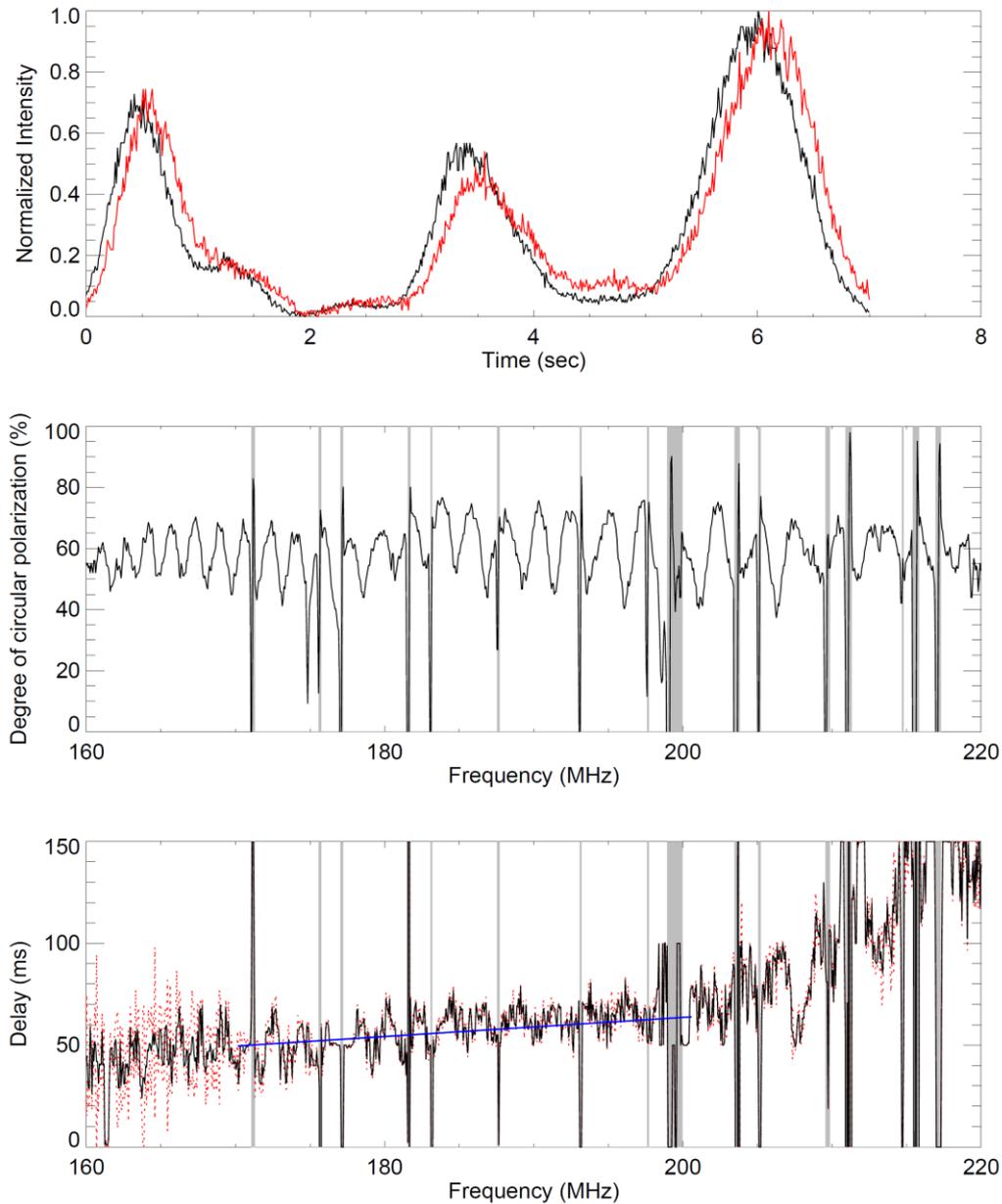

Figure 5

Polarization characteristics of ZP. Top: normalized light curves of the RCP (black line) and LCP (red line) components at 190 MHz (from 03:22:22 UT to 03:22:29 UT). Middle: frequency profile of the degree of circular polarization at the time of ZP enhancement (03:22:17 UT). Bottom: frequency dependence of the delay between the LCP and RCP components. The ordinate represents the maximum lag $l$ used in equation (3). The red dotted line denotes the error in the calculation of the delay estimated from equation (4), and the blue line is the fitted line in the frequency range of 170–200 MHz. Gray hatched regions are the frequency bands of artificial interferences.